\documentclass[11pt]{iopart}

\usepackage{iopams}  
\usepackage{graphicx}

\begin{document}

\title[Dust in the Interplanetary Medium]{Dust in the Interplanetary Medium}

\author{Ingrid Mann$^1$, Andrzej Czechowski$^2$, Nicole Meyer-Vernet$^3$, Arnaud Zaslavsky$^3$ Herve Lamy$^1$}

\address{$^1$ Belgian Institute for Space Aeronomy, Brussels, Belgium\\
$^2$   Space Research Center, Polish Academy of Sciences, Warsaw, Poland\\
$^3$ LESIA, Observatoire de Paris, Meudon, France}

\ead{ingrid.mann@aeronomie.be}

\begin{abstract}
The mass density of dust particles that form from asteroids and comets
in the interplanetary medium of the solar system is, near 1AU, comparable
to the mass density of the solar wind.
It is mainly contained in particles of micrometer size and larger. 
Dust and larger objects are destroyed by collisions and sublimation
and hence feed heavy ions into the solar wind and the solar corona.
Small dust particles are present in large number and as
a result of their large charge to mass ratio deflected by electromagnetic
forces in the solar wind. For nano dust particles of sizes $\simeq 1 - 10$ nm, 
recent calculations show trapping near the Sun and outside from about 0.15 AU
ejection with velocities close to solar wind velocity.
The fluxes of ejected nano dust are detected near 1AU with the plasma wave
instrument onboard the STEREO spacecraft.
Though such electric signals have been observed during dust impacts
before, the interpretation depends on several different parameters 
and data analysis is still in progress.
\end{abstract}


\section{Introduction}
Dust is an important component of space plasmas and the properties of 
dust and plasma are coupled through different processes. They include
the charging of the dust particles, the sputtering of the dust species 
by impinging plasma particles, the dynamical friction, and the 
electromagnetic forces.
The presence of dust particles in the 
plasma can lead to heating, cooling, and mass loading or unloading. Most of 
our understanding of the dust plasma interactions
in space is derived from theory. Laboratory measurements provide
only limited information. This is particularly true for the nano 
dust (of size \ensuremath{\le} 100 nm, mass \ensuremath{\le} 10$^{-18} kg$) 
which also evades most 
of the standard observational techniques. Dust particles reside 
in the solar wind that fills the interplanetary medium between the orbits
of the planets in the solar system. Dust and 
plasma in-situ measurements from spacecraft provide direct information 
and allow a close view of the dust and its interactions in the solar wind. 

In section (2) we will give an overview of the solar system dust cloud
and the different observations of dust. In section (3) we describe the 
different interactions of dust in the solar wind.
We then discuss  the dynamics and fluxes of nano dust in the solar wind 
(section 4) and end with a short summary (section 5). 

\section{Dust Origin, Evolution and Observations } 

\begin{figure}[ht]
\centering
\includegraphics[width=100mm,clip]{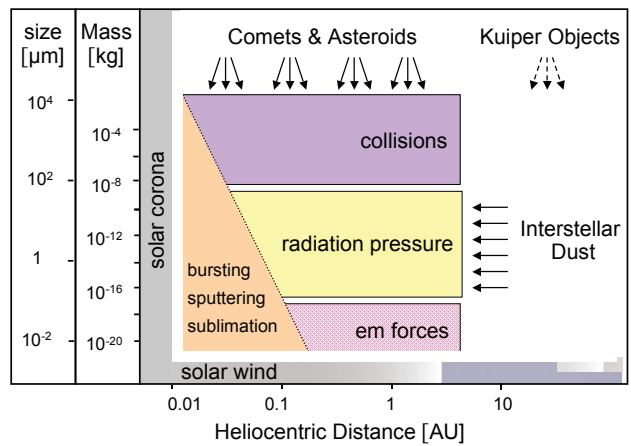}
\caption{\protect \label{overview_figure}
Sources and sinks of the solar system dust cloud.}
\label{fig:overview_figure}
\end{figure}

\paragraph{Origin} 
The vast majority of dust particles in the solar system form as a result of 
fragmentation of the small solar system objects (Fig. \ref{fig:overview_figure}).
Asteroids and comets, as well as meteoroids that originate from them,
are the major dust source inward of the orbits of the
terrestrial planets. 
Dust particles, except for the smallest ones, 
also enter the solar system from
the interstellar medium and these interstellar dust particles
form the most prominent dust component in the region of the orbits
of the giant planets and probably also beyond 
this region, where no in-situ measurements have been made.

\paragraph{Zodiacal light observations}

The dust particles produce the Zodiacal light  \cite{Leinert98}, which is made of two kinds of radiation. The first is scattering of sunlight, which makes the visible and near-infrared part. The second is thermal emission, which dominates the spectrum at longer wavelengths.
The Zodiacal light is the dominant unresolved brightness component of the night sky beyond the terrestrial atmosphere from visible to far infrared $\simeq100$$\mu $m, and it does not exhibit any temporal variation. Its brightness increases toward the Sun, with a local maximum extending along the ecliptic plane that contains the orbits of the planets. Local enhancements in brightness reveal regions of enhanced dust production along cometary orbits (cometary dust trails) and regions of dust accumulation related to asteroid fragments (dust bands). Cosmic dust is also directly observed in the vicinity of comets (dust tails).

\paragraph{Collected samples} 

Samples of cosmic dust are collected in the upper Earth atmosphere and analyzed with standard laboratory techniques \cite{Rietm02}.
Various measurements suggest that the dust contains a large fraction of silicates, metal oxides, sulfur compounds as well as carbon bearing species. Its elemental composition is close to that of the solar photosphere, but the dust is depleted in hydrogen and other light elements and there is little information
on the noble gases content. The dust particles have irregular shape and heterogeneous composition and a fraction of them is porous. The subunits out of which the heterogeneous dust particles are made cover
a broad range of sizes with average  \ensuremath{\approx} 100 $\mu $m. Nanometric sub-units are also observed and interplanetary dust particles frequently contain, for instance, FeNi particles of 5 - 10 nm size \cite{messenger}. At these small scales, analysis is difficult since full laboratory analysis requires samples of mass
$\geq10^{-12}$ kg (sizes $\gtrsim$ several $\mu$m) \cite{Zolensky2000}.

\paragraph{Dust flux near 1 AU} 

The  size distribution of solar system objects ranges from the kilometre size of comet nuclei and asteroids to dust smaller than micrometric size. It is is only known near  $1$AU from the Sun (the Earth's orbit). Meteor observations, lunar crater
studies and in-situ measurements with conventional dust detectors
reach down to approximate sizes of $0.1 \mu $m. Recent detections
of smaller dust will be discussed below, but the transition to molecule and atom sizes is outside the observing domain. The cumulative flux, $F(m)$ of particles with mass $\geq m$ per area and time unit follows roughly $F(m)\sim m^{-5/6}$. The flux is enhanced compared to this power law
in certain mass intervals (cf \cite{grun85}, \cite{Ceplecha1998}). The mass density of
solid objects of mass $m\leq10^{-3}$ kg at 1 AU is comparable to the solar wind mass density $\sim10^{-20}$kg$\cdot $m$^{-3}$ and is mainly provided by dust of size $s\sim 10\mu $m. These large dust particles also provide the maximum surface area (for interaction with the
solar wind and for producing thermal emission and scattered light).

\paragraph{Dust orbits and distribution } 

The motion of the dust in the solar system is determined by the initial
conditions (velocity and distance from the Sun) and by the acting forces.
The main  forces are solar gravity, radiation pressure,
solar wind pressure and the Lorentz force. The dust particles of size $\gtrsim$ 1$\mu $m
are predominantly influenced by the gravitational force, so that they move approximately in Keplerian orbits. 
Their initial orbits are close to that of the parent body. 
The perturbations modify the  perihelion of the orbits and the longitudes of the ascending node. 
In result, the dust cloud is approximately cylindrically symmetric relative to
the axis through the Sun perpendicular to the ecliptic. These large dust particles produce the vast majority of the observed Zodiacal light. Due to the non-radial
component of the radiation pressure force the particles gradually
lose orbital energy and angular momentum; this reduces the  semi-major axis and eccentricity of the orbits (i.e. Poynting-Robertson effect). As a result, the particles migrate toward the Sun on time scales $\leq10{}^{7}$ years. Note, that the impinging solar wind particles also transfer momentum onto the dust.
This process is generally negligible, but the tangential component of the momentum
transfer gives rise to the plasma Poynting Robertson effect.
It is of the same order of magnitude as the normal Poynting Robertson effect,
and more important in the early solar system as well as in some  planetary systems surrounding other stars \cite{minato2006}. During their passage near the Sun the dust
particles experience catastrophic collisions so that the
inner solar system dust cloud contains a large fraction of freshly
produced fragments. The Lorentz force plays an important role in the dynamics of the small grains, as described below. The interstellar dust particles stream into the inner solar system
from a direction similar to that of the interstellar wind \cite{Mann2010}. Those interstellar dust particles which are not deflected by the Lorentz force or the radiation pressure are focused
by the solar gravitational attraction.  

\paragraph{Dust detection} 

In-situ dust measurements in space are mainly based on the detection of  the
charges  generated by dust impacts (cf. \cite{Auer2001}).
Dust-generated charges also
influence plasma measurements from spacecraft, like ion mass spectrometer data
and magnetic and electric field measurements. 
The latter was observed for the first time in space aboard Voyager during Saturn planetary ring crossings ( \cite {Aubier1983} \cite {Gurnett1983}). Electric signals produced by dust impacts have also been studied during comet encounters and
other ring crossings (cf.\cite{Meyer-Vernet2001}). 
Since 2007 the STEREO/WAVES instrument \cite{bougeret08}
measures near 1AU a large number of voltage pulses of shape similar to that previously observed during dust impacts. The large number of events can be
explained by impacts of nano dust particles that are accelerated to large
velocities and hence generate a large impact charge \cite{meyer-vernet09a}.
Analysis of similar events measured onboard Cassini during Jupiter flyby
showed that they were simultaneous with the detection of nano dust particles by a conventional dust analyzer instrument \cite{meyer-vernet09b}.
Note that the conditions for dust charging and dynamics in planetary magnetospheres are different from those in the solar wind.

\section{Dust-generated species in the solar wind}

\paragraph{Dust effects on the solar wind} Dust particles are destroyed by mutual collisions, sublimation and sputtering, and  these processes feed the dust material into the solar wind in the form
of smaller solid particles, electrons, and atoms and molecules.
Since the solar wind mass resides in the H and He species, while the dust mass resides
in the heavier species, the destruction of dust particles can influence the
minor solar wind components. Several space measurements reveal phenomena that possibly result
from the presence of dust in the solar wind.
The Solar Wind Ion Composition Spectrometer (SWICS) aboard the Ulysses spacecraft discovered ions that originate from the vicinity of the Sun, are singly charged and contain the abundant elements as well as molecular species (inner source pick-up ions \cite{gloeckler10}.
Though  other possible explanations were given to explain the inner source pick-up ions,
model calculations of the collision rates in the interplanetary dust cloud within 1 AU suggest that dust destruction by mutual collisions generates ion fluxes that are comparable to the singly charged heavy ion fluxes of this inner source \cite{MannCzechowski2005}.
Dust collisions can also account for the observed molecular ions. 
Even so, the calculated ion fluxes vary by orders of magnitude depending
on model assumptions of the dust distribution. 
The measured ion fluxes were derived from a small number of events
accumulated over a wide range of spacecraft orbits. 
Other examples are the sporadic enhancements in interplanetary magnetic field  and enhanced fluxes of the neutral solar wind. 
Both phenomena are possibly related to the solar wind crossing the interplanetary dust cloud, but are not quantitatively described
yet \cite{mann_etal2010}.
Though the resulting fluxes of heavy ions are smaller than the heavy ions production
by dust destruction \cite{MannCzechowski2005},
some solar wind species may change their charge and energy due to interaction
with the solar wind. 
Impinging solar wind particles pass through the grain and pick electrons by charge-exchange with the atoms of the grain material and typically emerge neutralized or in a singly charged state, if they are not decelerated to a stop within a depth of several 10 nm.

\paragraph{Estimating dust sublimation}
Most of the dust particles
sublimate within 0.1 AU (about 20 solar radii) from the Sun. A canonical
value of a "dust free zone" radius around the Sun is about 4 solar radii, but this
cannot be derived from observations, and calculations show that some
dust particles with small absorptivity can reach a distance of about
2 solar radii \cite{mann2004}.
We briefly address
here sublimation of larger objects, since observations of sungrazing
comets prove that these objects survive very close to the Sun. 
Figure \ref{fig:sublimation}  shows the mass loss of large objects
falling into the Sun. 
The mass loss rates are calculated using ablation models similar to 
those used for the Earth's atmosphere \cite{CampbellKoschny2004}
and for Venus atmosphere \cite{mcauliffe2006}.
The calculations start with initial conditions at 20 solar radii and solve 
iteratively the equations for energy balance, mass loss, and deceleration.
This first-order model is one-dimensional
and assumes that objects fall straight to the Sun with a free-fall velocity. 
The left panel of Fig. \ref{fig:sublimation} shows that most of the mass of 
the objects sublimates within a very small range of radial distance. 
It also illustrates that ablation is differentiating in the sense that more refractory
materials can dive deeper into the corona.
The right panel shows the final distance (remaining mass $10^{-10}$ kg) 
of penetration of objects as a function of 
their initial mass. Objects with masses larger than a few tons can reach the
solar surface. 
This model allows estimating the mass deposition profile in the solar corona,
 as well as the influence of sungrazing comets in the corona 
(as does an improved version that is currently under development \cite{Lamy10}.  

\begin{figure}
\centering
\includegraphics[width=0.5\textwidth]{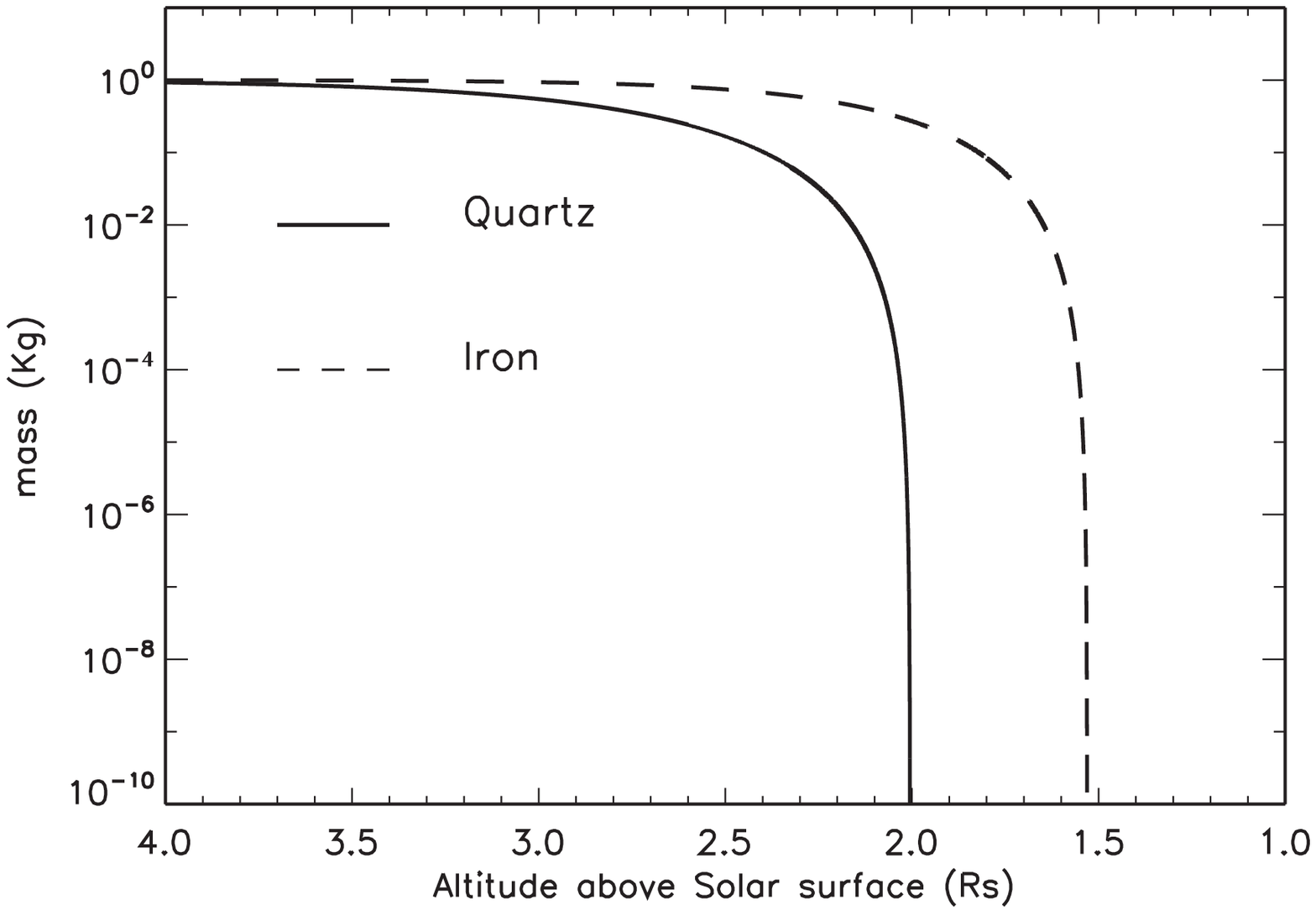}\includegraphics[width=0.475\textwidth]{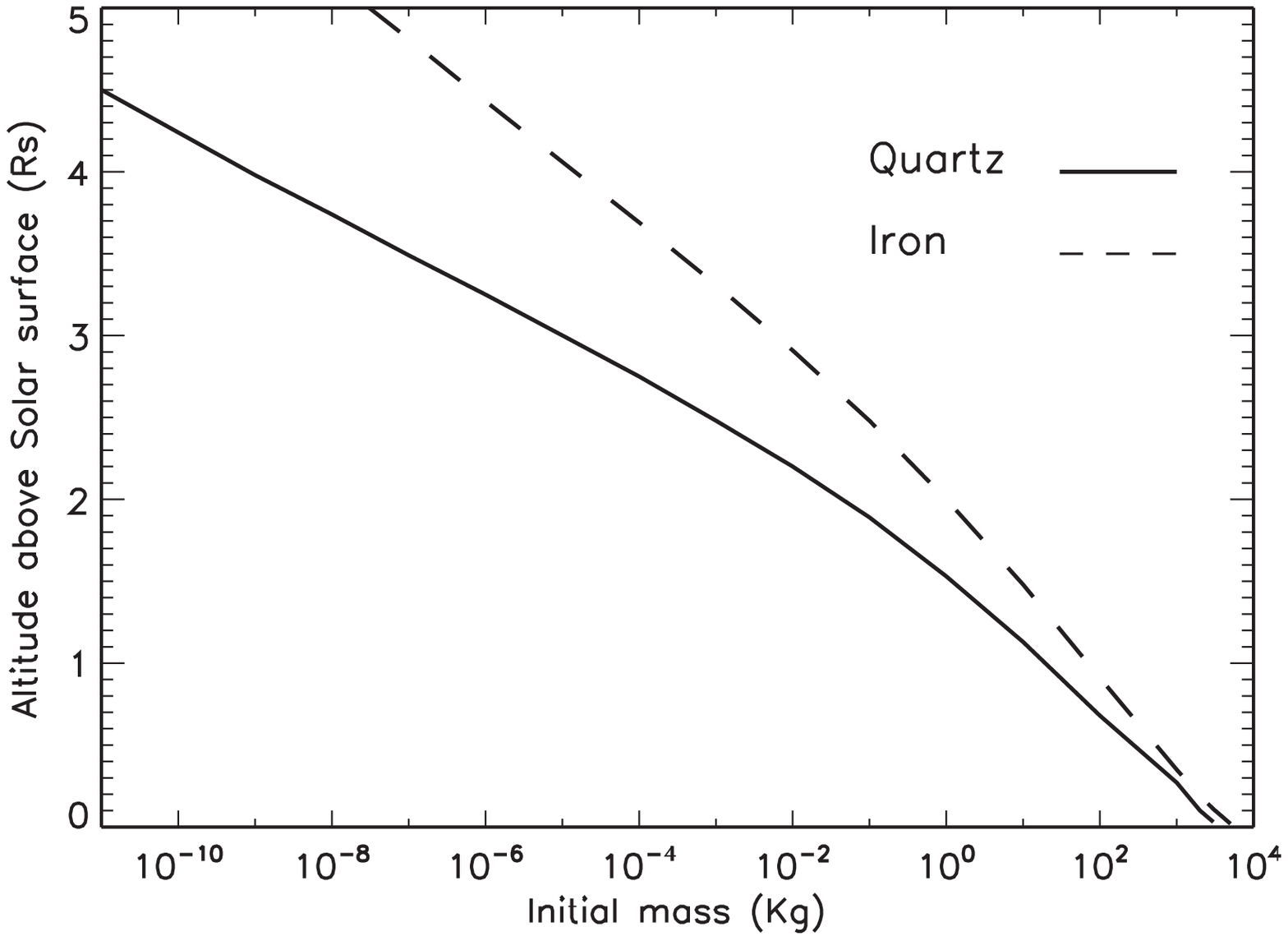} 
\caption{\protect \label{fig:sublimation}
Left: Mass versus altitude above solar surface for two objects with 1 Kg initial mass falling into the Sun.
Right: final altitude as function of the initial mass of the object \protect \cite{Lamy10}.}
\label{fig:sublimation}
\end{figure}

\paragraph{Estimating dust destruction by high velocity impacts}

High velocity impact of a dust particle on a solid target in space causes destruction, sublimation and ionisation of dust and target material.
This occurs during mutual collisions of
dust particles as well as during impact  on a spacecraft. The impacts are described empirically  by experiments carried out in accelerator facilities where the impact
generated charge is measured for grains of several 10 micrometer and
impact speeds of several 10 km/s. The charge, Q, that is produced increases steeply with the impact speed as \cite{mcbride1999}

\begin{equation}
Q \simeq 0.7 m^{1.02} v^{3.48}
\end{equation}

\noindent where charge $Q$ is in Coulombs, mass $m$ is in kg and impact speed $v$ is in km/s.
In another approach the vapour production is estimated via
the fragmentation laws for solids. The laws are based on the consideration of larger objects and macroscopic crater formation as a result of the
propagation of impact induced shock waves in bulk solids. The fragment distributions that are derived using this approach
match with the results of macroscopic impact experiments in size
ranges of mm and cm and impact velocities smaller than km/s.
An extrapolation was made for describing dust collisions in the interstellar medium \cite{jones94}.
These models predict vapour production for collision velocities of the order of km/s, but do
not consider the charge state. A similar approach simulating the shock wave propagation in the solid leads to the predictions of temperature $10^{4} K$ and neutral and low charge states for the species 
in the vapour cloud \cite{hornungkissel1994}.
 Both, the charge state and the temperature of the impact generated cloudlets are estimated with great uncertainty.

\paragraph{Estimating dust-generated species in the solar wind}
While the processes of dust destruction are qualitatively known, they are not well quantified, which,
together with the lack of knowledge of absolute fluxes of dust and larger objects near
the Sun, prevents an exact estimate of the production rates.  For example, the calculated vapour production from dust collisions
within 1 AU varies between $2\cdot10^{2}$kg$\cdot $s$^{-1}$ and $4\cdot10^{4}$kg$\cdot $s$^{-1}$
depending on the assumed dust distribution and dust material composition \cite{MannCzechowski2005}
and due to the large number of unkown parameters
no reasonable estimate can be given for the mass that dust sublimation feeds into the solar
wind and into the solar corona at distances $\leq0.1$ AU. The species that form by dust destruction are assumed to be in majority initially neutral or singly charged. The main ionization mechanisms
are photoionization and charge exchange with solar wind protons.
These processes ionize an oxygen or carbon atom at 1 AU within $10^{6}s-10^{7}s$, which is less than the dust orbital period. The formed singly ionized species are picked up within the Larmor gyration time of the order of seconds, and species that are picked up outward of 0.1 AU are likely to keep
their acquired charge state \cite{MannCzechowski2005}.
The lifetimes of the neutrals decrease rapidly toward the Sun, where electron
impact ionization becomes important and the ions are multiply charged.
The species that are released in the inner corona, finally, have similar charge states
as the solar wind ions.

\section{Dust trajectories in the solar wind and fluxes near 1 AU}

\paragraph{Detection of nano dust}

\begin{figure}
\centering
\includegraphics[width=100mm,clip]{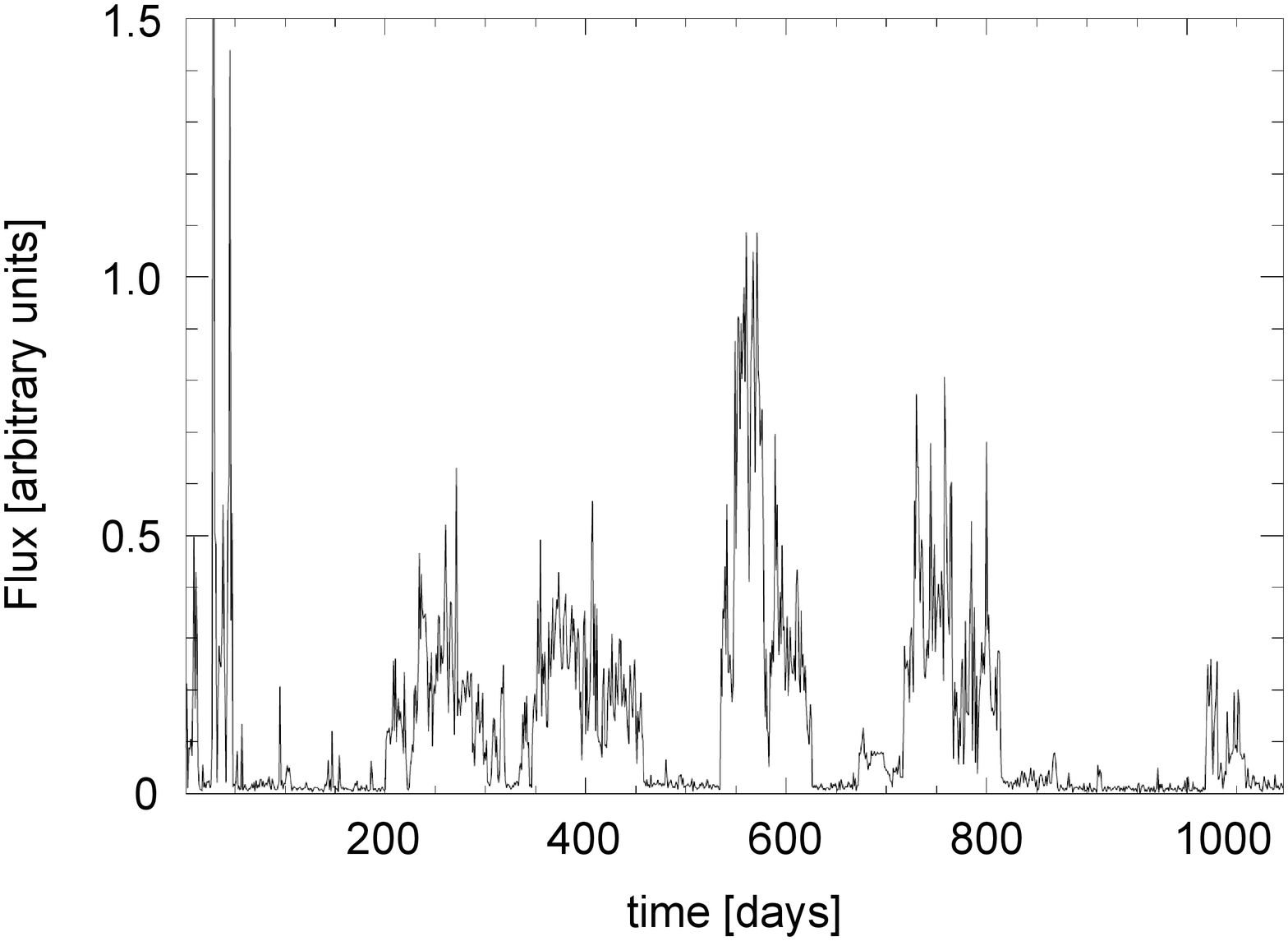} \caption{\protect \label{flux}
The flux rate of nano dust measured with the plasma wave experiment S/WAVES onboard
Stereo A from 2007 to 2009  \protect\cite{Zaslavsky10}.}
\label{fig:flux}
\end{figure}

The nano dust detected onboard STEREO has a high flux with large time fluctuations \cite{meyer-vernet09a}.
Figure \ref{flux} shows the variation in nano dust
impact rate during the last 3 years  \cite{Zaslavsky10}.
Understanding the data requires studying the trajectories of small
charged dust particles and estimating their fluxes.
Since fragmentation rates increase toward the Sun, the nano dust  may
form  by collisional fragmentation of larger dust particles  moving  in bound Keplerian orbits inward of 1 AU.

\paragraph{Dust in the solar wind}

The size of the dust particles is small compared to the solar wind Debye length
$\sim 10$ m and gyration (Larmor) radius ($\sim 10^{3}$ m for
electrons at 1 AU). The solar wind flux onto the dust can be
considered as single particle impacts at the solar wind
speed ($\sim 300-800$ km s$^{-1}$) for ions and the thermal speed for
electrons ($\sim 10^{3}$ km  s$^{-1}$). 
The dust particles are mainly charged by collection of solar wind
particles and photoionization. 
The latter process is generally dominant and leads to a positive surface potential 
of the order of several Volts on a time scale $\tau\propto1/a$ (for a dust particle of  
size $a\approx1\mu$m at 1 AU, $\tau \approx 1$s).
In the interplanetary medium (except very close to the
Sun where secondary electron emission gains importance) the
dust potential does not vary  with distance, nor with grain size \cite{kimura98},
so that the ratio of surface charge to mass $q/m \propto1/ a^{2}$.
The charging process
has no noticeable influence on the surrounding solar wind plasma
and conditions for dusty plasma collective effects are not fulfilled.
Since  $q/m$ determines the coupling of the grains to the magnetic field, this coupling is larger for smaller grains. 

\paragraph{Dust trajectory calculations}

In this subsection we discuss the results of the first study of the dynamics 
of the nano dust particles in the interplanetary medium dust cloud that were carried 
out recently \cite{czechowski10}. 
All previous studies were concerned with larger 
grains, with much lower surface charge to mass ratios. 
The equations of motion are solved  assuming  initial circular orbits at  the  release distance  $r_0$, taking into account the  Lorentz
force, solar gravity  and radiation pressure (the latter term is found to be unimportant for the nanodust
because of small value of the radiation pressure to gravity ratio of 0.1). The magnetic field follows a Parker
spiral, with regions of opposite polarity separated by a current
sheet. 
The solar wind speed is 400 km/s near the solar equator and 800 km/s at high latitudes. 
The calculations were made for constant charge-to-mass 
ratios $ q/m=10^{-7} e/m_p$ to $10^{-4} e/m_p$. 
This corresponds to an approximate size of 100 nm for the former, 
and of 3 nm for the latter $q/m$.
A general finding of the calculations is that there exists a trapping region
in the vicinity of the Sun, extending to about 0.15 AU in the ecliptic and 
farther at higher latitudes. 
The nano dust released inside this region is trapped in bound orbits, while
the nano dust released at larger distances is accelerated outward. 
The trapped particles  move in non Keplerian orbits which can reach very low perihelion distances.  
The ejected particles roughly follow an outward trajectory that stays close to the rotating magnetic field line. 
Their gyro-frequency decreases quickly with increasing 
distance from the Sun.  The orbits of the grains with
$ q/m=10^{-4}- 10^{-5} e/m_p$ are found to be rather similar, 
with a gyration radius that is smaller for the smaller dust (larger $q/m$).

\paragraph{Qualitative picture of dust trajectories}

\begin{figure}[ht]
\centering
\includegraphics[width=0.5\textwidth]{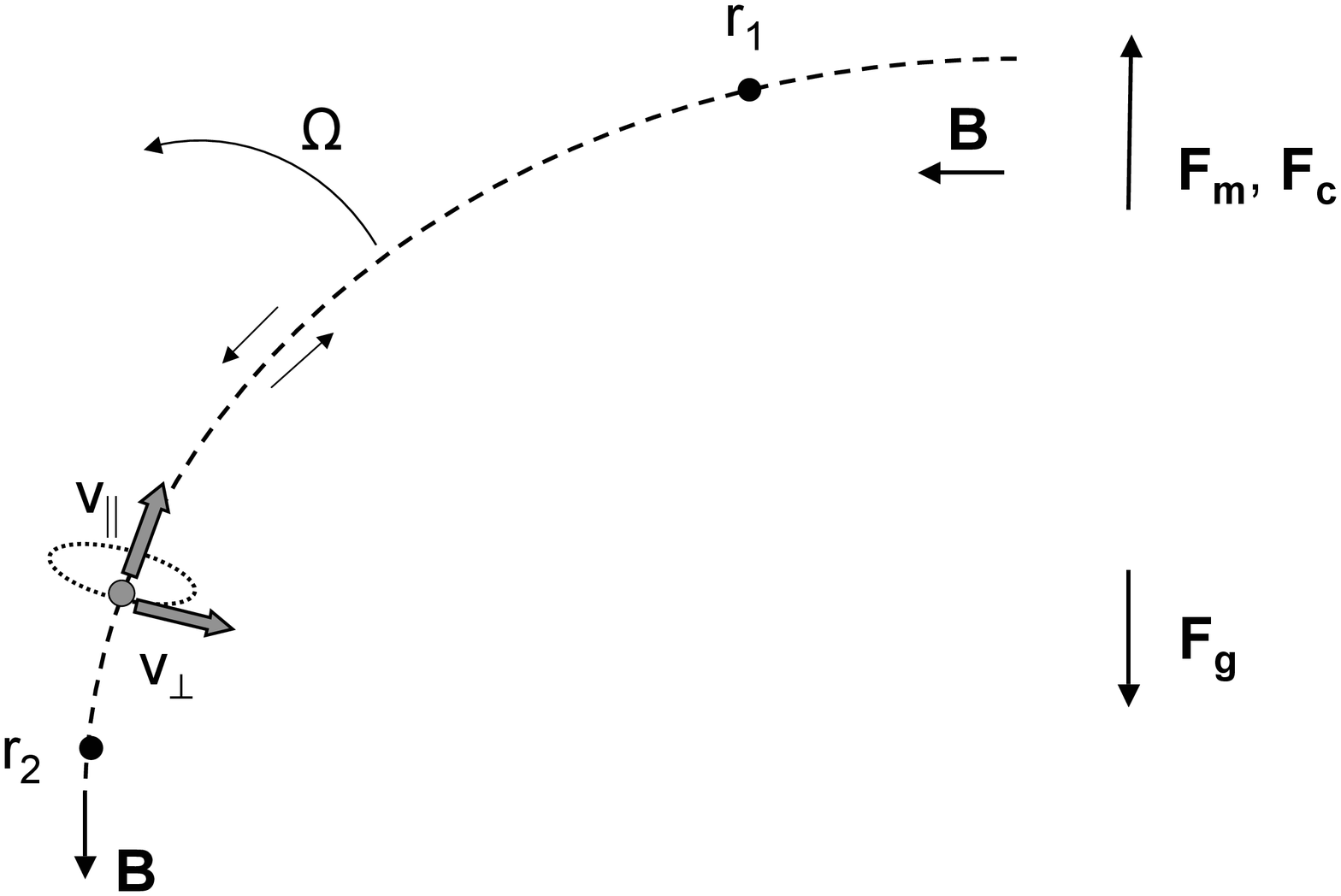}\includegraphics[width=0.5\textwidth]{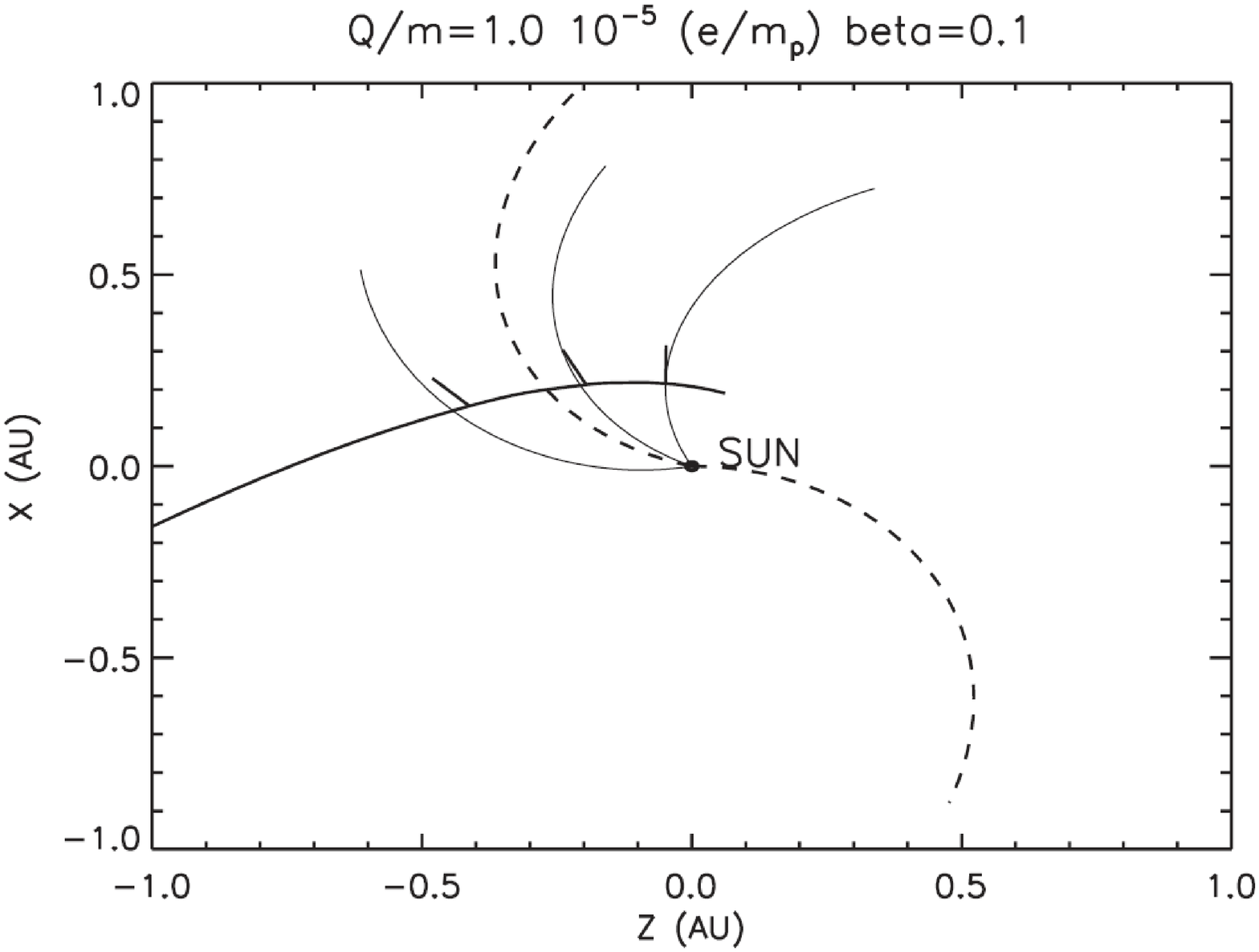} 
\caption{\protect \label{orbits_sketch}
The left side depicts the motion in the 
guiding center approximation for particles that are trapped along a rotating 
magnetic field line between position $r_{1}$ and  $r_{2}$. 
The right hand side shows a calculated trajectory 
for  the  dust particle with  
$q/m=10^{-5}e/m_{p}$ that is released from a circular orbit of radius 0.2 AU close
to the solar equator plane (shown projected
onto this plane \protect \cite{czechowski10}). }
\label{fig:orbits_sketch}
\end{figure}

Although the reported calculations didn't use the guiding centre approximation,
it helps to understand the nanodust dynamics 
qualitatively for both trapped and ejected particles 
(\cite{czechowski10}, see Fig. \ref{orbits_sketch}). 
The guiding center approximation decomposes the 
equation of motion into a guiding center component following the magnetic field
line, components describing the gyration about the guiding center and 
the drift component.   
This  approximation contains terms that can be 
interpreted as  the  magnetic mirror force, solar gravity and  the  
centrifugal force. The dust particle gyrates about the magnetic field 
line with  the  velocity ${\bf {V}_{\perp}}$. 
It follows the rotation of the magnetic field
line and (in the case of a trapped particle) oscillates along it. 
The motion along the field
line is governed by the solar gravity $F_{g}$ directed inward, and
the centrifugal force $F_{c}$ and the  mirror  force $F_{m}$, that are 
pointing outward. The inner radius of the  orbit  is proportional  
to  $F_{m}/$$F_{g}$, whereas the outer is 
proportional to $F_{g}/F_{c}$. Particles are ejected
when the combination of centrifugal and magnetic mirror force exceeds 
gravity.
Note that the actual trajectories
drift away from the moving field lines. 
This is shown on the right of
Fig. \ref{orbits_sketch} for one particular dust trajectory,
but looks similar for other cases. 
The magnetic field line that passes through the point
of release at the time of release is shown at two subsequent
moments of time and the nano dust positions at these moments are marked.
The dashed line depicts the current sheet  intersection with the 
ecliptic at the time of release.

\noindent Calculations agree with the qualitative discussion based on guiding 
center approximation. The maximum radius
of  the  trapping  region  varies for dust released from orbits with 
different inclinations and it is different for  the case 
 when  the particles are released close to the solar equator
and for the case  when  they are released at the point of maximum 
latitude of the orbit. The trapping region extends further outward for 
the dust released a high latitudes, but dust densities
are small there. 
The perihelia of trapped orbits are in the range 
$0.002AU\leq r_{perihelion}\leq0.032AU$
and they are smallest for the dust released close to the outer
edge of the trapping zone. The time to reach the first perihelion is of the 
order of several days. 
Typical sublimation distances of dust are
$r\simeq0.025AU$ and, even though the scattering and thermal properties
that determine the sublimation of the nano dust are unknown, it is safe to assume that
most of  the trapped dust is quickly destroyed.
The smallest dust with $q/m=10^{-4}- 10^{-5} e/m_p$ 
released outside the trapping limit quickly gains velocity and reaches 
speeds close to 300 km/s at 1AU. 
The results of orbital calculations  assuming fixed charge  do not 
depend strongly on $q/m$ for $q/m \geq 10^{-5}$.
This suggests 
that, despite the small (several elementary charges) average 
charge of the small grains,  charge fluctuations  may  not play a 
significant role.

\paragraph{The influence of the current sheet}

\begin{figure}
\centering
\includegraphics[width=0.5\textwidth]{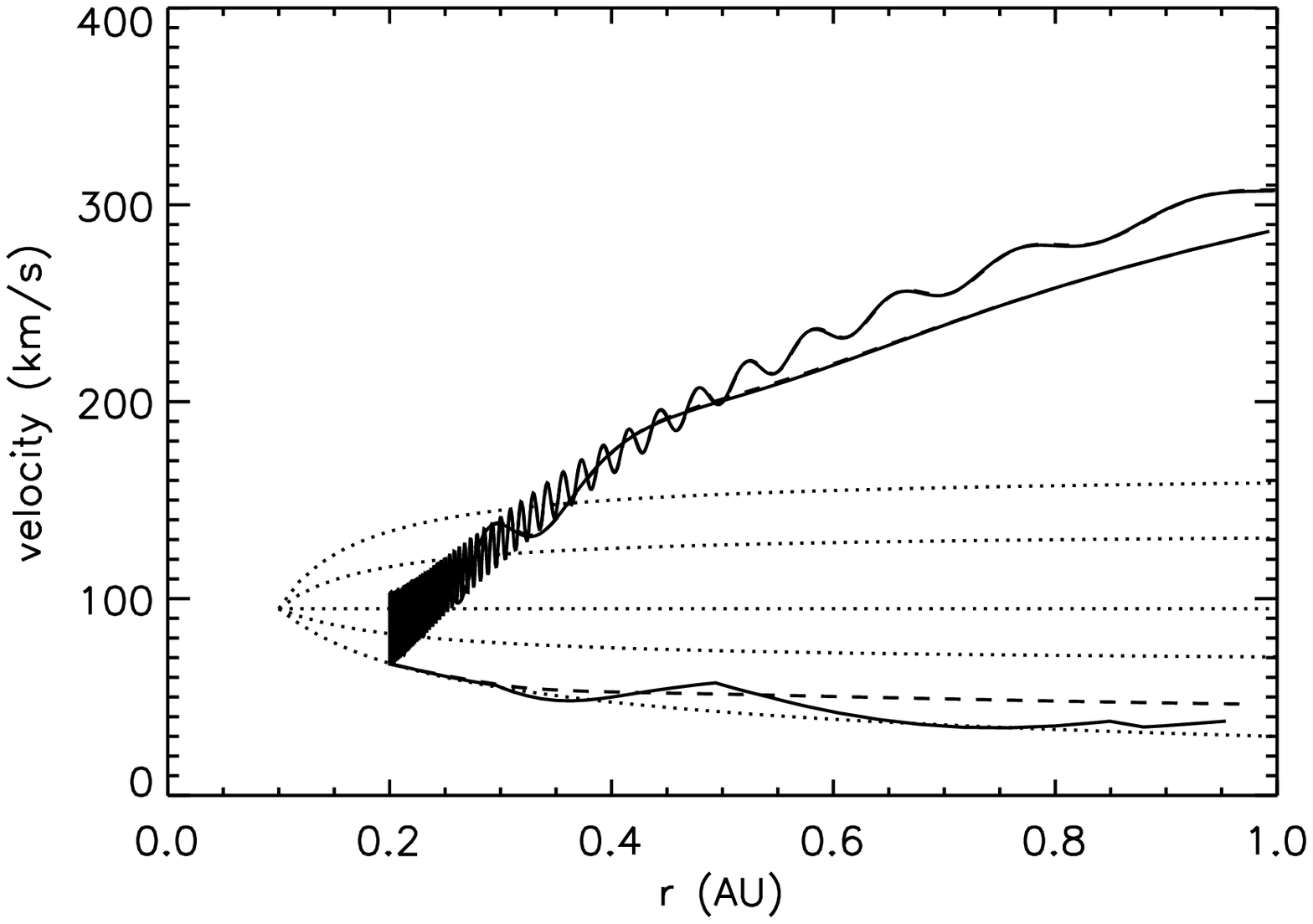}\includegraphics[width=0.5\textwidth]{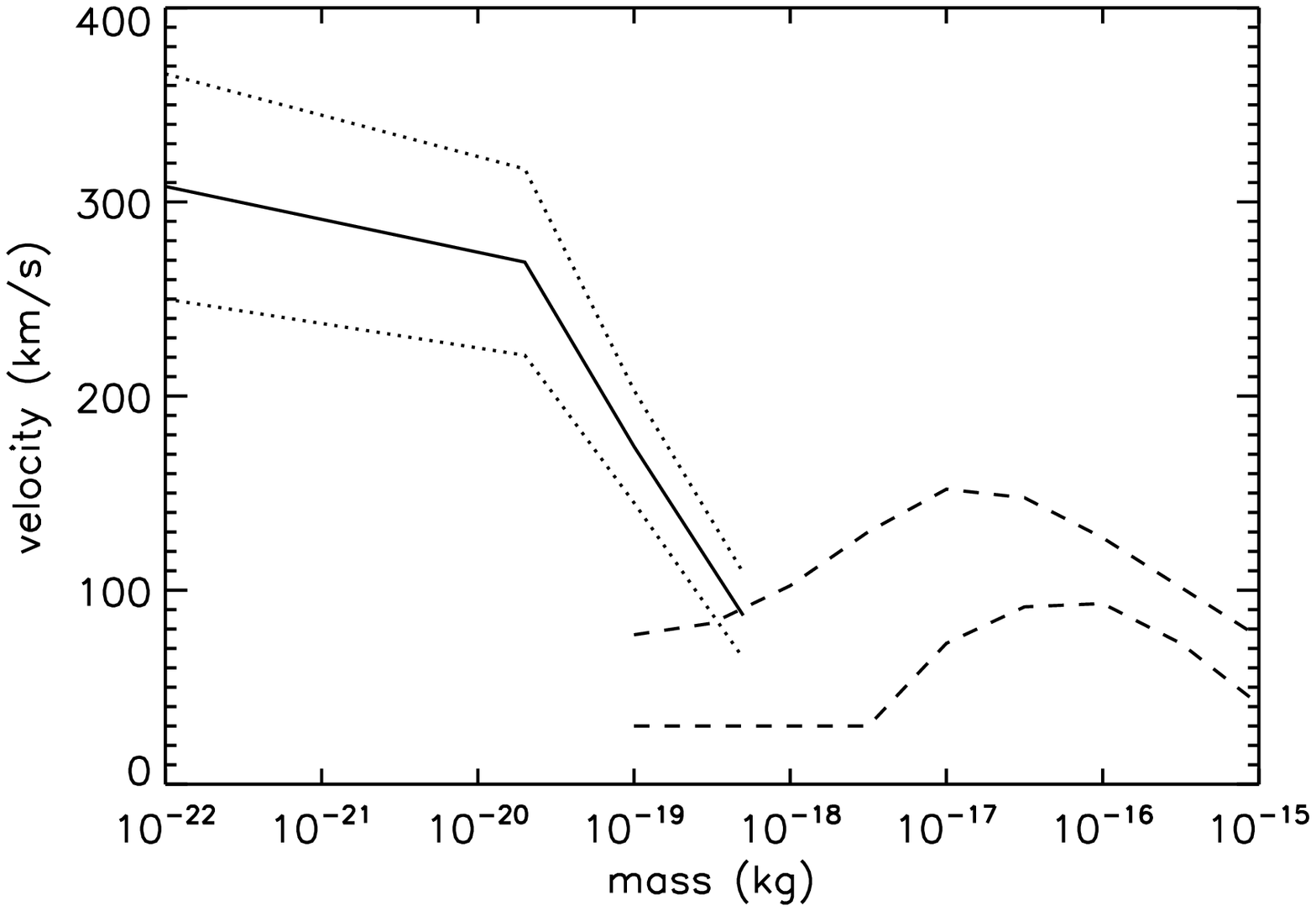} \caption{\protect \label{Veldistance}
Speed vs. distance from Sun for dust ejected by electromagnetic forces (solid and dashed lines) 
from top to bottom with $ q/m=10^{-4},10^{-5},10^{-7} $$e/m_{p}$. 
The dotted lines  show velocities of dust ejected by radiation pressure 
with from top to bottom: $F_{r}/F_{g}=$ 2, 1.5, 1, 0.75, 0.5.
The right panel shows the average velocity as function of mass at 1 AU for dust ejected by electromagnetic forces (solid line and dotted lines) and dust ejected by radiation pressure
(dashed lines) based on the models described in the text.}
\label{fig:Veldistance}
\end{figure}

Figure \ref{fig:Veldistance}  shows absolute velocities vs. distance
from the Sun for  the  grains with different $q/m$.
During field configuration with incoming magnetic field in
the northern hemisphere (when the electric field
-(1/c) V$\times$B points toward the current sheet), the 
speeds drop during current sheet crossing and then increase again. 
This is shown with the solid lines. 
Current sheet crossing does not occur  frequently during the
opposite field configuration with outgoing magnetic field in
the northern hemisphere and electric field pointing away
from the current sheet (shown with dashed lines). 
For $q/m=10^{-4}e/m_{p}$ and $ q/m=10^{-5}e/m_{p}$  the
shown curves for the different field configurations overlap. 
Dust acceleration quickly drops down for smaller dust charges and 
grains with $q/m$=10$^{-7}$ (about 100 nm) are ejected with moderate speed. 
During the phase of the solar cycle when the electric field
points away from the current sheet these larger particles 
are ejected toward the solar poles.

\paragraph{Dust ejection by electromagnetic forces compared to radiation pressure}
In Figure \ref{fig:Veldistance} we compare the dust ejected by electromagnetic force
with the larger dust ejected by the radiation pressure force
(so-called $\beta$-meteoroids, note that in this manuscript 
$\beta = F_{r}/F_{g}$ denotes the radiation pressure to gravity ratio).
The left panel shows the velocity, $v$ at distance, $r$ from the Sun 
of the nano dust (from \cite{czechowski10}) and of the $\beta$-meteoroids calculated
(with gravity constant, G  and mass of the Sun, $M_{o}$) from 

\begin{equation} 
v(r_{0},r)=
\sqrt{\frac{GM_{o}}{r}}\cdot\left(2-2\beta
+\frac{\left(2\beta-1\right)}{r_{0}}\cdot r\right)^{1/2}
\end{equation}

\noindent assuming release of $\beta$-meteoroids from circular orbit of 
radius  $r_{0}$ = 0.1 AU, since this is the region of largest collision rates. 
Dust that is affected by radiation pressure is ejected outward for $\beta\geq 0.5$ 
whereas for $\beta\geq 1$ velocities increase with distance, but final velocities are far below
those of the nano dust.
The right panel shows average velocities vs. mass at 1 AU.
The shown values for the nano dust  
are derived based on a model of nano dust production by collisional 
fragmentation inside 1 AU combined with the reported trajectory calculations \cite{czechowski10}. 
The $\beta$-meteoroids velocities are derived by applying the radiation pressure to gravity ratios 
vs. mass obtained for a highly absorbing cometary dust model (upper curve) and  a 
dirty silicate asteroidal dust model (lower curve).
The latter is the more common case within the solar system cloud
and the values for both cases are taken from \cite{Wilck96}.
The curves  show the  absolute values  of the velocity, but  
in all considered cases the trajectories have significant 
radial velocity components. 
This comparison shows that the dust ejected by electromagnetic
forces and the  $\beta$-meteoroids have distinctly different velocities.

\paragraph{Model calculations in comparison to STEREO observations}
The trajectory calculations show that nanodust particles move at high speeds near 1 AU, which explains why they can be detected on STEREO. Applying dust collision models with several assumptions provides a flux estimate $\phi(a>10$nm)$\simeq 0.6$m$^{-2}$s$^{-1}$ which agrees with the STEREO fluxes. This yields nanodust mass and momentum fluxes much smaller than the solar wind ones.  The observed nanodust flux variations are not yet explained. The collision model shows that the majority of the nano dust is generated by large dust particles in bound orbits, whose flux is unlikely to vary on the short time scales observed. Similarly, since the dust is accelerated in the inner solar system, a correlation of the flux rate with the magnetic field measured onboard STEREO is unlikely. However, since about $ 80-90 \% $ of the nano dust is trapped near the Sun, small variations of the extension of the trapping zone may produce the flux variations observed.  Variations in magnetic field and solar wind near the Sun can also cause flux variations.

\section{Summary}

We have briefly summarized the different contributions of interplanetary dust
to the solar wind. The processes that release dust material
into the solar wind appear relatively simple, but the absolute fluxes are not
known. Since most dust destruction occurs near the Sun, future space
missions into the inner solar system could possibly reveal the influence
of the dust generated species on the minor solar wind ions.
The acceleration of nano dust is a newly found interaction with the solar wind.
The detection of nano dust fluxes with plasma wave measurements
is a new result that requires further studies of the detection
process as well as of the different parameters that influence the
dust speed and absolute fluxes. Likewise, the formation of nano dust during
collisional fragmentation of larger dust, its composition and properties
need further studies.

\ack
This paper was prepared while I.M. stayed as Visiting Astronomer at LESIA, CNRS, Observatoire de Paris Meudon and as Visiting Professor at Universite Diderot, Paris and this support is greatly acknowledged. This work was partly supported by the German Aerospace Center DLR (project RD-RX /50QP 0403), the Belgium Solar Terrestrial Center of Excellence and the Polish Ministry of Science (grant N N203 4159 33).

\end{document}